\documentclass[12pt]{article}\newcommand{\figw}{0.9}

\usepackage{amsmath}
\usepackage{graphicx}

\newcommand{\eql}[1]{\label{#1}}
\newcommand{\eq}[1]{(\ref{#1})}
\newcommand{\figl}[1]{\label{fig#1}}
\newcommand{\fig}[1]{\ref{fig#1}}
\newcommand{\average}[1]{\langle #1\rangle}
\renewcommand{\vec}[1]{{\bf #1}}
\let\oldauthor=\author
\newsavebox{\theau}
\renewcommand{\author}[1]{\sbox{\theau}{\large #1}}
\newcommand{\affiliation}[1]{\oldauthor{\usebox{\theau}\\\normalsize\em 
#1}}
\providecommand{\pacs}[1]{\ \ \ PACS: #1}
\newcommand{\paper}{paper}
\newcommand{\Eq}[1]{equation~\eq{#1}}
\newcommand{\Eqs}[1]{equations~\eq{#1}}
\newcommand{\Fig}[1]{figure~\fig{#1}}
\newcommand{\numkopje}[1]{\section{#1}}
\newcommand{\subkopje}[1]{\subsection{#1}}
\newcommand{\kopje}[1]{\section*{#1}}
\newcommand{\Ref}[1]{reference~\cite{#1}}
\let\olddate=\date
\renewcommand{\date}[1]{\olddate{#1}\maketitle 
\newcommand{\maketitle}{}}
\newcommand{\primenumber}{\renewcommand{\theequation}{\arabic{equation}$'$}\addtocounter{equation}{-1}}
\newcommand{\unprimenumber}{\renewcommand{\theequation}{\arabic{equation}}}
\newcommand{\inline}[1]{\primenumber\begin{equation}#1\end{equation}\unprimenumber}
\newcommand{\pinline}[1]{\[#1\]}

\begin{document}

\title{An Extension of the Fluctuation Theorem}
\author{R.~van~Zon and E.G.D.~Cohen}
\affiliation{{\ }\vspace{-5mm}\\\em The Rockefeller University\\
\em 1230 York Avenue, New York, NY 10021}
\date{May 7, 2003}

\begin{abstract}
Heat fluctuations are studied in a dissipative system with both
mechanical {\em and} stochastic components for a simple model: a
Brownian particle dragged through water by a moving potential.  An
extended stationary state fluctuation theorem is derived.  For
infinite time, this reduces to the conventional fluctuation theorem
only for small fluctuations; for large fluctuations, it gives a much
larger ratio of the probabilities of the particle to absorb rather
than supply heat.  This persists for finite times and should be
observable in experiments similar to a recent one of Wang~{\em et~al}.
\end{abstract}


\maketitle

\numkopje{Introduction}

There is a lack of unifying principles in nonequilibrium statistical
mechanics, compared to the equilibrium case. So it is not surprising
that the {\em fluctuation theorem} has received a lot of attention, as
it gives a property of fluctuations of entropy production for a large
class of systems, possibly arbitrarily far from
equilibrium\cite{Evansetal93,Evansetal94,GallavottiCohen95a,%
Kurchan98,LebowitzSpohn99,CohenGallavotti99,%
Maes99,Wangetal02,ReyBelletThomas02,VanZonCohen02b}.

In terms of heat rather than entropy production, the conventional
stationary state fluctuation theorem (SSFT)\footnote{We call \Eq{E1}
the ``conventional SSFT'' to distinguish it from the extended SSFT
presented in this \paper.}
states that the probability $P_\tau(Q_\tau)$ to find a value of
$Q_\tau$ for the amount of heat dissipated in a time interval $\tau$,
satisfies, in a nonequilibrium stationary
state\cite{Evansetal93,GallavottiCohen95a},
\begin{equation}
   \frac{P_\tau(Q_\tau)}{P_\tau(-Q_\tau)} \sim  e^{\beta Q_\tau},
\eql{E1}
\end{equation}
where $\sim$ indicates the behavior for large $\tau$.  Here
$\beta=(k_BT)^{-1}$, with $k_B$ Boltzmann's constant and $T$ the
(effective) temperature of the system. In contrast, the transient
fluctuation theorem (TFT) considers fluctuations $Q_\tau$ in time,
when the system is initially in equilibrium\cite{Evansetal94}.  These
theorems were first demonstrated in deterministic many particle
systems in an external field
\cite{Evansetal93,Evansetal94,GallavottiCohen95a}, but later in
stochastic systems as well\cite{Kurchan98,LebowitzSpohn99}.

In this context, it is customary to identify $Q_\tau/T$ as a
(generalized) entropy production.  Then, \Eq{E1} is interpreted as a
theorem for entropy production arbitrarily far from equilibrium.
Furthermore, the fluctuation theorem holds for arbitrary values of
$Q_\tau$, i.e., also far from its average. Hence, it is often referred
to as a large deviation theorem.

Recently, a laboratory experiment was carried out by Wang {\em et
al.}\cite{Wangetal02}. They measured fluctuations in the work done on
a system in a transient state of a Brownian particle in
water, subject to a moving, confining potential. The TFT was
confirmed.

In the deterministic models, dissipation is often modeled by including
a damping term in the equations of motion, chosen such as to keep,
e.g., the total energy of the system fixed. This damping term is a
mechanical expression for what in reality is an external thermostat.
Any work done by the external field on the system is absorbed by this
``internal'' thermostat.  So the external work done on the system
equals the heat dissipated in the system.

While Wang {\em et al.} intended to study the entropy production (or
heat) fluctuations, in fact, the work fluctuations were
studied\cite{VanZonCohen02b}. In contrast to the above sketched purely
deterministic models, work fluctuations differ from heat fluctuations
in their system due to the presence of a confining potential. Thus,
some of the external work done is converted into potential energy and
only the rest is converted into heat.  In fact, in this system the
work fluctuations in the stationary state satisfy the conventional
SSFT\cite{VanZonCohen02b,MazonkaJarzynski99}, but the heat
fluctuations do not, as we shall show.  As it turns out, having a {\em
deterministic} together with a {\em stochastic} motion results in a
behavior of the heat fluctuations that coincides with the conventional
SSFT only for a restricted set of small fluctuations and in a very
different behavior for larger ones.

\numkopje{Work fluctuations -- conventional SSFT}

We first discuss the work-related fluctuation theorem in the
experiment of Wang {\em et al.}, which was treated theoretically in
Refs.~\cite{VanZonCohen02b,MazonkaJarzynski99,Sevicketal02}.  The
Brownian particle in a fluid, subject to a harmonic potential moving
with constant velocity $\vec v^*$, was described by an overdamped Langevin
equation:
\begin{equation}
  \frac{d\vec x_t }{dt} = - (\vec x_t-\vec x_t^*) + \zeta_t.
\eql{E2}
\end{equation}
Here, $\vec x_t$ is the position of the particle at time $t$, $\vec
x_t^*=\vec v^* t$ is the position of the minimum of the harmonic
potential at time $t$ [cf. \Eq{E7P}], and $\zeta_t$ is a fluctuating
force with zero mean and a delta function correlation in time. We
remark that the relaxation time of the position of the particle has
been set equal to one. Also, we set $k_BT=1$, so $\average{\zeta_t\zeta_s} =
2\delta(t-s)$.\footnote{Compared to Ref.~\cite{VanZonCohen02b},
$k_BT=1$, the harmonic force constant $k=1$ and the friction
coefficient $\alpha=1$.  So the energy unit is $k_BT$, the time unit
is the relaxation time $\alpha/k$, and the length unit is the thermal
width $\sqrt{k_BT/k}$.} In \Ref{VanZonCohen02b}, it was shown that
\Eq{E2} is solvable in a co-moving frame, in which it reduces to a
standard Ornstein-Uhlenbeck process.  Thus, the stationary probability
distribution and Green's function are known, and are both Gaussian in
$\vec x_t$.  The work is the total amount of energy put into the
system in a time $\tau$. This is a fluctuating quantity, given
by\footnote{A force on the particle implies an opposite force on $U$,
which must be overcome by an external force to keep $U$ moving. The
work done by this force is $W_\tau$ in \Eq{E3}\cite{VanZonCohen02b}.}
\begin{equation}
  W_\tau \equiv \vec v^*\cdot\int_0^\tau [- (\vec x_t-\vec x^*_t)]dt.
\eql{E3}
\end{equation}
Here, the time $t=0$ denotes the initial time of an interval of length
$\tau$ in the stationary state.  $W_\tau$ is a linear function of the
positions $\vec x_t$, and since those have a Gaussian probability
distribution function, so does $W_\tau$.  When the mean and variance
of the probability distribution function $P^W_\tau$ are computed
[using the stationary solution and Green's function of \Eq{E2}], one
finds\cite{VanZonCohen02b}
\begin{equation}
  \lim_{\tau\rightarrow\infty} \frac{1}{w\tau}
        \ln\left[\frac{P_\tau^W(pw\tau)}{P_\tau^W(-pw\tau)}\right] 
  = p.
\eql{E4}
\end{equation}
Here, $p$ is a scaled value of $W_\tau$, defined as
$p=W_\tau/\average{W_\tau}$, such that $\average{p}=1$. We also wrote 
\begin{equation} 
  \average{W_\tau}=w\tau,
\eql{E5}
\end{equation}
with $w$ the average work rate, 
which is independent of $\tau$ in
the stationary state.  In the current units, $w = |\vec v^*|^2$.  
Equation \eq{E4} is, for the work fluctuations, a
more careful formulation of the SSFT in \Eq{E1}.  A work related TFT
also holds\cite{VanZonCohen02b,MazonkaJarzynski99,Sevicketal02}.

\numkopje{Heat fluctuations}

\subkopje{Fourier transform of the distribution}

We now turn to the heat SSFT. The heat $Q_\tau$ is that part of the
work $W_\tau$ that goes into the fluid. Some work is also stored in
the potential, so
\begin{equation}
  Q_\tau \equiv  W_\tau - \Delta U_\tau,
\eql{E6}
\end{equation}
where $\Delta U_\tau$ is the change in potential energy of the
particle in a time $\tau$,
\begin{equation}
  \Delta U_\tau \equiv  U_\tau - U_0,
\eql{E7}
\end{equation}
with \inline{U_t \equiv \frac12 |\vec x_t - \vec x^*_t|^2.\eql{E7P}}
This form of $U_t$ makes $Q_\tau$ nonlinear in $\vec x_t$.  As a
result, the probability distribution function $P_\tau(Q_\tau)$ of
$Q_\tau$ need not be Gaussian. Nonetheless, it is possible to compute
its Fourier transform.

The Fourier transform of $P_\tau(Q_\tau)$, defined as
\begin{equation}
  \hat P_\tau(q)\equiv
 \int_{-\infty}^\infty\!dQ_\tau\,e^{iqQ_\tau}P_\tau(Q_\tau),
\eql{E8}
\end{equation}
is computed by writing $P_\tau$ as [using \Eqs{E6} and \eq{E7}]
\begin{equation}
  P_\tau(Q_\tau) = \iint\! d{\vec x}_0 \, d{\vec x}_\tau \,
  P_\tau^{W_\tau,\vec x_0,\vec x_\tau}
  (Q_\tau+\Delta U_\tau,\vec x_0,\vec x_\tau),
\eql{E9}
\end{equation}
where $P_\tau^{W_\tau,\vec x_0,\vec x_\tau}$ is the joint distribution of
the work $W_\tau$, the positions $\vec x_0$ and $\vec x_\tau$ at the
beginning and at the end of the time interval $\tau$,
respectively. This distribution is Gaussian because $W_\tau$, $\vec x_0$,
and $\vec x_\tau$ are all linear in $\vec x_t$. When \Eq{E9} is
inserted into \Eq{E8}, a seven dimensional Gaussian integral is left,
which after some algebra yields
\begin{equation}
  \hat P_\tau(q) = \frac{
			\exp\left\{ w q (i-q)\left[\tau - 
	\frac{2q^2(1-e^{-\tau})^2}{1+(1-e^{-2\tau})q^2}\right]\right\}
			}
			{\left[1+(1-e^{-2\tau})q^2\right]^{3/2}}.
\eql{E10}
\end{equation}

Once $\hat P_\tau(q)$ has been transformed back, one considers
\begin{equation}
  f_\tau(p) \equiv \frac{1}{w\tau}\ln\left[
	\frac{P_\tau(pw\tau)}{P_\tau(-pw\tau)}\right].
\eql{E11}
\end{equation}
Here, $p$ is a scaled value of $Q_\tau$, defined as
\pinline{p=Q_\tau/\average{Q_\tau},} i.e., $\average{p}=1$. We also used that
\pinline{\average{Q_\tau}=\average{W_\tau}-\average{\Delta U_\tau}=w\tau} by
\Eq{E5}, since $\average{\Delta U_\tau}=0$ in the stationary state.  The
scaled logarithmic ratio $f_\tau(p)$ should be equal to $p$ for
$\tau\rightarrow\infty$ when the conventional SSFT holds.

As far as we know, there is no exact result for the inverse Fourier
transform of $\hat P_\tau(q)$ in \Eq{E10} in terms of known functions.
Therefore, a completely analytic treatment did not seem
feasible. Instead, we used first a numerical method, the fast Fourier
transform algorithm\cite{NumericalRecipes}, to invert \Eq{E10}. The
resulting probability distribution function $P_\tau$ as well as the
corresponding $f_\tau$ have been plotted in \Fig{1}. These results do
not agree very well with the straight line with slope one, which
should be approached for large $\tau$ if the conventional SSFT were to
hold.  One might think that this is due to $\tau$ not being large
enough.  However, we found that deviations of $f_\tau(p)$ from $p$ for
large $p$ are generic, while the straight line {\em is} approached
only for smaller $p$.  Nonetheless, we cannot say anything
conclusive about the large $\tau$, large $p$ behavior because the
distribution gets very peaked and hence becomes smaller for large
deviations, which makes the numerical method unreliable.

\begin{figure}[t]
\centerline{\includegraphics[width=\figw\textwidth]{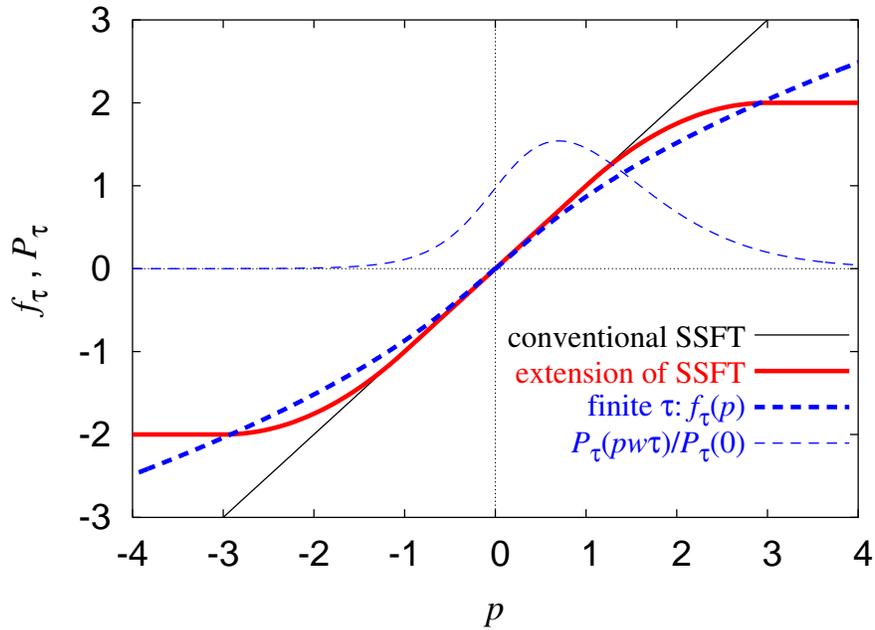}}
\caption{Numerically obtained $f_\tau(p)$ (bold dashed) for $v^*=1.5$
and $\tau=1.3$, versus the ($v^*$ independent) extension of the SSFT for
$\tau\rightarrow\infty$ (bold solid).  Also plotted are the
conventional SSFT (thin solid), and the numerically obtained
distribution function $P_\tau(pw\tau)$, scaled by its value
at zero (thin dashed).}\figl{1}
\end{figure}

\subkopje{Large deviation method -- extended SSFT}

Therefore, we used next an analytical asymptotic approach based on
large deviation theory\cite{Spohnbook} similar to the treatment by
Lebowitz and Spohn\cite{LebowitzSpohn99}. One considers then
\begin{equation}
  e(\lambda) \equiv \lim_{\tau\rightarrow\infty} -\frac{1}{w\tau}
		     \ln \average{e^{-\lambda Q_\tau}}.
\eql{E12}
\end{equation}
This infinite-$\tau$ quantity is used to reconstruct the
distribution function of $Q_\tau$ for large $\tau$ by setting
\begin{equation}
  P_\tau(Q_\tau) \sim \exp\left[-w\tau \hat e(Q_\tau/w\tau)\right],
\eql{E13}
\end{equation}
where $\hat e(p)$ is the Legendre transform of $e(\lambda)$:
\begin{equation}
  \hat e(p) = \max_{\lambda} [e(\lambda)- \lambda p].
\eql{E14}
\end{equation}
For a class of models, Lebowitz and Spohn
proved the symmetry relation
\begin{equation}
  e(\lambda) = e(1-\lambda).
\eql{E15}
\end{equation}
From this, using \Eqs{E11}, \eq{E13} and \eq{E14}, one sees that
$\lim_{\tau\rightarrow\infty} f_\tau(p)=p$, i.e., the conventional
SSFT holds\cite{LebowitzSpohn99}.

Our numerical results suggest, however, that for our model the
conventional SSFT for the heat does not hold.  We therefore expect
\Eq{E15} to be violated.  Indeed, the following calculation of
$e(\lambda)$ shows this to be the case.

The Fourier transform $\hat P_\tau(q)$ in \Eq{E10} determines
$e(\lambda)$. First, from \Eq{E8}, we have
\begin{equation}
  \average{e^{-\lambda Q_\tau}}
	\equiv \int_{-\infty}^{\infty} \!dQ\,e^{-\lambda Q_\tau}P_\tau(Q_\tau)
	=  \hat P_\tau(i\lambda) 
\eql{E16},
\end{equation}
Thus we need the analytic continuation of $\hat P_\tau$ to imaginary
arguments.  This poses no difficulty as long as $\hat
P_\tau$ remains analytic. One finds from \Eqs{E10} and \eq{E16}
\begin{equation}
 \average{e^{-\lambda Q_\tau}}
  =\frac{\exp\left[-w\lambda(1-\lambda)\left\{
	\tau+\frac{2\lambda^2(1-e^{-\tau})^2}{1-(1-e^{-2\tau})\lambda^2}
  \right\}\right]}{\left[1-(1-e^{-2\tau})\lambda^2\right]^{3/2}}.
\!
\eql{E17}
\end{equation}
Clearly, there are divergences at the singular points $\lambda=\pm
(1-e^{-2\tau})^{-1/2}$, where the right-hand side (r.h.s.) of \Eq{E16} is no
longer analytic, so that \Eq{E17} only holds for values of $\lambda$ in
between those. Using \Eqs{E12} and \eq{E17}, we have
\begin{equation}
  e(\lambda) = \lambda (1-\lambda)\quad\mbox{for $|\lambda|<1$},
\eql{E18}
\end{equation}
where, taking $\tau\rightarrow\infty$ as in \Eq{E12} moves the
singularities to $\pm 1$. This $e(\lambda)$ satisfies
\Eq{E15} for $0<\lambda<1$.

However, as $\lambda$ approaches the singularities, the function in
\Eq{E17} diverges. Beyond the singularities at
$\pm(1-e^{-2\tau})^{-1/2}$, the r.h.s. of \Eq{E17} becomes purely {\em
imaginary}, and multivalued due to the denominator. But the {\em
left-hand side} of \Eq{E17} remains {\em real}. Clearly, we cannot use
\Eq{E17} for $|\lambda|>(1-e^{-2\tau})^{-1/2}$.  To determine
$\average{e^{-\lambda Q}}$ in that case, we first need to know why the
integral in \Eq{E16} diverges as $\lambda\rightarrow\pm
(1-e^{-2\tau})^{-1/2}$.  As we will argue next, this happens because
$P_\tau$ has exponential tails.  Since $P_\tau(Q_\tau)$ is a
normalized distribution and $e^{-\lambda Q_\tau}$ a regular function,
the divergence in \Eq{E16} can only be due to the behavior of the
integrand at $\pm\infty$. In fact, for $\lambda>0$, any divergence
must be due to the behavior at negative $Q_\tau$ and for $\lambda<0$,
it must be due to the behavior at positive $Q_\tau$.  Now, for
$\lambda<0$, if the distribution function $P_\tau(Q_\tau)$ fell off
faster than exponential for large positive $Q_\tau$, the factor
$e^{-\lambda Q_\tau}$ could not make the integral diverge. But as it
does diverge, we conclude that the distribution function falls off
exponentially or slower. On the other hand, if it did fall off slower
than exponential, then the exponential factor $e^{-\lambda Q_\tau}$
would always dominate the distribution function for large positive
$Q_\tau$ and the integral in \Eq{E16} would diverge for all
$\lambda<0$. Since there {\em are} negative values of $\lambda$ for
which the integral converges, the function $P_\tau$ cannot fall off
slower than exponential. Hence, it must fall off exponential for large
$Q_\tau$. Considering $\lambda>0$, one deduces along similar lines
that it must fall off exponentially for large negative values of
$Q_\tau$ as well.

In fact, the integral in \eq{E16} diverges for all $|\lambda|\geq
(1-e^{-2\tau})^{-1/2}$.  For if the function $P_\tau(Q_\tau)$ falls off
exponentially for large positive $Q_\tau$, say as $e^{-aQ_\tau}$, the
integral in \eq{E16} diverges for all $\lambda\leq-a$. Likewise, given
that the $P_\tau(Q_\tau)$ falls off like $e^{aQ_\tau}$ for large negative
$Q_\tau$, the integral diverges for all $\lambda\geq a$.  Hence, for
$|\lambda|\geq (1-e^{-2\tau})^{-1/2}$, the quantity on the r.h.s of
\Eq{E12} of which the limit is taken, is minus infinity for all $\tau$,
so that $e(\lambda)=-\infty$. Thus, \Eq{E18} becomes
\begin{equation}
  e(\lambda) = \left\{
		\begin{array}{ll}
			\lambda(1-\lambda) 
			&\mbox{for $|\lambda|<1$}\\
			-\infty 
			&\mbox{otherwise.}
		\end{array}
		\right.
\eql{E19}
\end{equation}
This $e(\lambda)$ does {\em not} satisfy the symmetry relation in
\Eq{E15}, e.g., for $\lambda=-1/2$, $e(\lambda)=-3/4$ whereas
$e(1-\lambda)=-\infty$. The fact that \Eq{E15} is not satisfied, means
that the conventional SSFT does not hold.  To make this more precise,
we use \Eqs{E14} and \eq{E19}, to find
\begin{equation}
  \hat e(p)
	=
	\left\{
	\begin{array}{ll}
	   -p & \mbox{for $p < -1$}\\
	   (p-1)^2/4 & \mbox{for $-1 \leq p\leq 3$}\\
	   p-2 & \mbox{for $p>3$.}	   
	\end{array}
	\right.
\eql{E20}
\end{equation}
Note that via \Eq{E13}, the large $|p|$ behavior is indeed
exponential\footnote{Singularities affected the distribution of a
different quantity as well in J.~Farago, J.\ Stat.\ Phys.\ {\bf 107},
781 (2002).}.  Using \Eqs{E11}, \eq{E13} and \eq{E20}, we find
\begin{equation}
  \lim_{\tau\rightarrow\infty} f_\tau(p) =
	\left\{
	\begin{array}{ll}
	p&\mbox{for $0\leq p < 1$}\\
	p-(p-1)^2/4&\mbox{for $1 \leq p < 3$}\\
	2&\mbox{for $p\geq 3$.}
	\end{array}
  \right.
\eql{E21}
\end{equation}
For negative $p$, we have \inline{f_\tau(-p)=-f_\tau(p).\eql{E21P}}
Equation \eq{E21} is an extension of the conventional SSFT. It
coincides with it for the middle region $-1<p<1$,\footnote{A
restriction to the SSFT (i.e., $|p|<1$) was also found in a model of
heat conduction in \Ref{ReyBelletThomas02}.} but differs from it for
other $p$ values. Most notably, for $p\geq3$, it attains a constant
value of two.

\subkopje{Saddle-point method -- large finite times}

If we compare the exact prediction of \Eq{E21} (plotted as the bold
solid line) with the numerical results (bold dashed line) in \Fig{1},
a clear discrepancy emerges: The curve of $f_\tau$ keeps increasing
with increasing $p$, whereas \Eq{E21} predicts that it should level off
to a value of two. This turns out to be a finite $\tau$ effect. To
prove this, we need a better treatment for large but not
infinite~$\tau$.  This can be obtained from a saddle-point method
applied to $e(\lambda)$, which we will present in a future
publication\cite{VanZonCohen03b}.  The saddle-point method gives
reliable results for sufficiently large $\tau$, as can be verified by
a comparison to our numerical results\cite{VanZonCohen03b}.  The
asymptotic behavior for large $\tau$ is then given by
\begin{equation}
  f_\tau(p) = \left\{ \begin{array}{ll}
                   p + h(p)/\tau  + O(\tau^{-2}) &\mbox{for $p<1$}
\\
             2+\sqrt{8(p-3)/\tau} + O(\tau^{-1})&\mbox{for $p>3$,}
             \end{array}\right.
\eql{E22}
\end{equation}
where \pinline{h(p)=\frac{8p}{9-p^2} - \frac{3}{2w}
\ln\left[\frac{(3-p)(1+p)}{(3+p)(1-p)}\right].} We left out the
behavior in between $p=1$ and $p=3$ as it does not appear to give any
additional insight. Equation \eq{E22} shows that as a function of $p$,
the function $f_\tau(p)$ increases like $\sim\sqrt{p-3}$ (for fixed
$\tau$), while as a function of $\tau$, it decreases (for fixed
$p$). As expected, for $\tau\rightarrow\infty$, it approaches the
large deviation result in \Eq{E21} as $1/\tau$ for small $p$ and as
$1/\sqrt{\tau}$ for large $p$.

Whether the new features beyond $p=1$ are observable depends on the
values of $P_\tau(pw\tau)$ and $P_\tau(-pw\tau)$. If these are too
small, the corresponding $f_\tau(p)$ cannot be seen in an experiment.
But the value of the distribution function as plotted in \Fig{1} is
non-negligible for values for $p$ (and $-p$) at which $f_\tau$
bends away from the conventional SSFT. Furthermore, the values
$v^*=1.5$ and $\tau=1.3$ used in \Fig{1} are realistic, as in the
experiment of Wang {\em et al.}  $v^*\approx 2.5$ and $\tau$ goes up
to $\approx6$. So this behavior should be experimentally detectable.

\numkopje{Discussion}

Summarizing, we have shown that the behavior of heat fluctuations in a
dissipative system with a {\em deterministic, mechanical} component
(the potential), and a {\em stochastic} component (the heat bath,
i.e., the water), differs from that known from both purely
deterministic systems and purely stochastic systems, in two
respects. a) For infinite $\tau$, the behavior of the conventional
SSFT is only seen for the scaled heat fluctuation $p$ between $-1$ and
$1$. For $p>1$, after a parabolic region between $p=1$ and $3$,
the quantity $f_\tau$ no longer increases, but stays at a plateau
value of two [for $p<-1$, see \Eq{E21P}].  b) The finite $\tau$
behavior of the conventional SSFT is in general unknown, but in our
case we find that $f_\tau$ keeps increasing with $p$. However
$f_\tau(p)$ stays well below the conventional SSFT , implying a larger
ratio of the probabilities of the particle to absorb rather than
supply heat.  These features are observable.

One of the striking features of the extended SSFT, the plateau value
of two for large (infinite) $\tau$ and large $p$, can be understood
physically.  For large $\tau$ and large $Q_\tau$, i.e. $p$, the
exponentially distributed $\Delta U_\tau$ far outweighs the Gaussian
distributed $W_\tau$ in \Eq{E6}.  The distribution of $\Delta U_\tau$
is exponential for large values ($\propto e^{-\beta|\Delta U_\tau|}$),
because it is the difference of the potential energies of the particle
at two times [cf.\ \Eq{E7}], which are both Boltzmann-like
distributed (because of the presence of the water) and independent of
each other for large $\tau$.  As $\average{Q_\tau}=w\tau$, this
leads to $P_\tau(Q_\tau)\propto e^{-\beta|Q_\tau-w\tau|}$, which
yields $P_\tau(Q_\tau)/P_\tau(-Q_\tau)\approx e^{2\beta w\tau}$ (if
$Q_\tau>0$). Then \Eq{E11} and $\beta=1$ give $f_\tau\approx 2$ of
\Eq{E21}.

We only considered an extension of the SSFT here. An extension of the
TFT can also be obtained. While for $\tau\rightarrow\infty$, one gets
again \Eq{E21}, for finite times, the extended TFT and SSFT
differ\cite{VanZonCohen03b}. Furthermore, the extended TFT differs
fundamentally from the conventional TFT, in that the latter holds as
an identity for all $\tau$, whereas the former one holds only in the
$\tau\rightarrow\infty$ limit\cite{CohenGallavotti99}.

Finally, one may wonder about the generality of the extended SSFT.
Our Langevin-based theory is only applicable near equilibrium. The
arguments given above suggest that the extended SSFT could hold for
other potentials as well. Perhaps it could even hold for a larger
class of systems, not near equilibrium, with mechanical and stochastic
components, since these are the main physical ingredients in our
theory.

\kopje{Acknowledgments}

We wish to thank M.J. Feigenbaum, F. Bonetto, and S. Ciliberto for
useful discussions.  This work was supported by the Office of
Basic Engineering of the US Department of Energy, under grant
No. DE-FG-02-88-ER13847.

\end{document}